\documentclass[aps,prb,twocolumn,groupedaddress]{revtex4-1}

\usepackage{etex}
\usepackage{amsmath}
\usepackage{courier}
\usepackage[pdftex]{graphicx}
\usepackage[table]{xcolor}
\usepackage{verbatim}
\usepackage{xfrac}

\begin{document}


\title{Perfect Anomalous Reflection with a Binary Huygens' Metasurface}


\author{Alex M.H. Wong and George V. Eleftheriades}
\email[]{gelefth@ece.utoronto.ca}
\affiliation{The Edward S. Rogers Department of Electrical and Computer Engineering, University of Toronto, Toronto, Canada M5S 3G4}


\date{\today}

\begin{abstract}
In this paper we propose a new metasurface that is able to reflect a known incoming electromagnetic wave into an arbitrary direction, with perfect power efficiency. This seemingly simple task, which we hereafter call perfect anomalous reflection, is actually highly non-trivial due to the differing wave impedances and complex interference between the incident and reflected waves. Heretofore, proposed metasurfaces which achieve perfect anomalous reflection require complicated, deeply subwavelength and/or multilayer element structures which allow them to couple to and from leaky and/or evanescent waves. In contrast, we demonstrate that using a Binary Huygens' Metasurface (BHM) --- a passive and lossless metasurface with only two cells per period --- perfect anomalous reflection can be achieved over a wide angular and frequency range. Through simulations and experiments at 24 GHz, we show that a properly designed BHM can anomalously reflect an incident electromagnetic wave from $\theta_i = 50^\circ$ to $\theta_r = -22.5^\circ$, with perfect power efficiency to within experimental precision.
\end{abstract}


\maketitle


\section{Introduction}
The ability to direct the flow of electromagnetic waves at will has captured the fascination of scientists and engineers. Potential applications include scientific and medical imaging, information and communication technology, and energy harvesting among many others. The recent decades have witnessed the emergence of metamaterials and metasurfaces which provide designer-defined properties to direct electromagnetic waves in arbitrary manner \cite{Veselago1968,Pendry2000,Shelby2001,Eleftheriades2002,Fang2005,Alu2005,Pendry2006,Jacob2006,Silveirinha2006,Smolyaninov2007,Landy2008,Gharghi2011,Sievenpiper1999,Holloway2005,Yu2011,Maci2011,Holloway2012,Pfeiffer2013PRL,Selvanayagam2013OptEx,Kim2014,Lin2014}. Specifically, a phase-gradient metasurface has been heavily investigated as a ubiquitous tool to control the reflection and transmission of a known incident wave \cite{Yu2011,Lin2014}. More recently, the Huygens' metasurface --- a surface which provides both electric and magnetic responses to an incoming EM wave --- has garnered attention as a metasurface with ultimate wave manipulation capabilities \cite{Pfeiffer2013PRL,Selvanayagam2013OptEx,Wong2014PNFA,Kim2014,Selvanayagam2014,Epstein2016NatComm}. For instance, it has been shown that total transmission and redirection of an incident wave is possible with a passive bianisotropic Huygens' metasurface, implementable on three closely-spaced layers of reactive elements \cite{Wong2016}.

The problem of arbitrarily \textit{reflecting} an electromagnetic wave, however, has proven surprisingly tricky. Prior to works on metasurfaces, researchers studying blazed gratings --- gratings which perform efficient retroreflection --- found that some thick gratings could redirect a wave into directions other than retroreflection with very high efficiency. These so-called ``off-Bragg'' gratings were numerically demonstrated for small to moderate separation angles \cite{Jull1979,Cho1997,Chen2013}. A theoretical investigation \cite{Maystre1981RadSci} found, from the perspective of plane wave diffraction, that perfect anomalous reflection was possible given the successful achievement of two angular parameters. The extent to which these parameters can be tuned remained unclear, but the authors of \cite{Maystre1981RadSci} further showed that for specific incidence and diffracted angles, the required parameters can be achieved by tuning the groove depth of a rectangular groove grating. However, recent works on metasurfaces \cite{Asadchy2016,Estakhri2016,Epstein2016PRL} considered the electromagnetic waves at an impedance boundary, and showed that, in order to steer an incident wave into an ``anomalous'' direction with perfect efficiency, one needs a reflection surface which is lossy (absorbs the incident wave) in some areas and active (reradiates more power than in the incident wave) in other areas. This finding raises questions on how a passive structure, like a grating, can achieve perfect anomalous reflection. Further, this explains why phase-gradient metasurfaces \cite{Yu2011} and some Huygens' metasurfaces \cite{Pfeiffer2013PRL,Selvanayagam2013OptEx} failed to achieve perfect anomalous reflection, but instead generated undesirable spurious reflection and/or transmission components.

Several alternatives have thus been proposed to achieve anomalous reflection. Ref. \cite{Kim2014} first reported the reduction of spurious reflection in a metasurface with an appropriately designed loss profile. Later, refs. \cite{Asadchy2016,Estakhri2016} showed that for the plane wave reflection case, one can completely eliminate spurious components by including loss in a passive metasurface. This allows one to achieve anomalous reflection at an efficiency limited by impedance mismatch, given by

\begin{equation}
\label{eq:ImpedanceMismatch}
\left( \frac{P_{out}}{P_{in}} \right) _{max} = \min \left( \frac{\cos\theta_r}{\cos\theta_i}, \frac{\cos\theta_i}{\cos\theta_r} \right). \\
\end{equation}

\noindent Alternatively, more power can be anomalously transmitted if one also allows some of the incident wave to be scattered into other directions. However, in both these cases, impedance mismatch makes it impossible to efficiently redirect a wave to or from near-grazing angles. More recently, refs. \cite{Epstein2016PRL,Diaz-Rubio2017,Asadchy2017ACSPhot} proposed specialized metasurfaces which redistribute power along the surface using auxiliary evanescent waves \cite{Epstein2016PRL} or leaky waves \cite{Diaz-Rubio2017,Asadchy2017ACSPhot}, to achieve the aforementioned absorptive and radiation regions using a passive and lossless metasurface. However, in these cases, the respective authors have not suggested a straightforward process through which one can design physical metasurface structures capable of achieving the aforementioned power transfer or auxiliary wave generation. Furthermore, the proposed metasurfaces are conceptually complex, in that they require very intricate coupling mechanisms, deeply subwavelength element dimensions and intricate metallization patterns and/or multilayer bianisotropic metasurface elements \cite{Epstein2016PRL,Diaz-Rubio2017,Asadchy2017ACSPhot}.

In this paper we report a very simple metasurface that successfully facilitates the aforementioned power redistribution and achieves perfect anomalous reflection, without any explicit involvement of surface or leaky waves. We show that maximally discretized metasurfaces --- with as few as two simple elements per period --- can be built to reflect an electromagnetic wave from a given input direction to a desired output direction. As an example, we demonstrate perfect anomalous reflection from an incident direction $\theta_{i} = 50^\circ$ to a reflection direction $\theta_{r} = -22.5^\circ$ using a Binary Huygens' Metasurface (BHM) --- a Huygens' metasurface with only two elements per period. Experimental results corroborate with simulation to show near-perfect power transfer from the incident to the desired reflection direction, and show that efficient reflection of greater than $90 \%$ power transfer from the incident to the redirected wave can be achieved over a wide bandwidth of $25.7 \%$. Our results demonstrate that such perfect anomalous reflection can be achieved without resorting to intricate, active or lossy metasurface profiles, and that this realization leads to superior metasurfaces capable of broadband performance. Furthermore, the development in this paper reconciles current works on metasurfaces with previous results in dielectric and metallic gratings: it offers a plausible explanation to the fundamental electromagnetics behind previous studies of near-perfect anomalous reflection from off-Bragg gratings.

\section{Results}
\subsection{Aggressive Metasurface Discretization}

\begin{figure}[t]
\includegraphics[width=85mm]{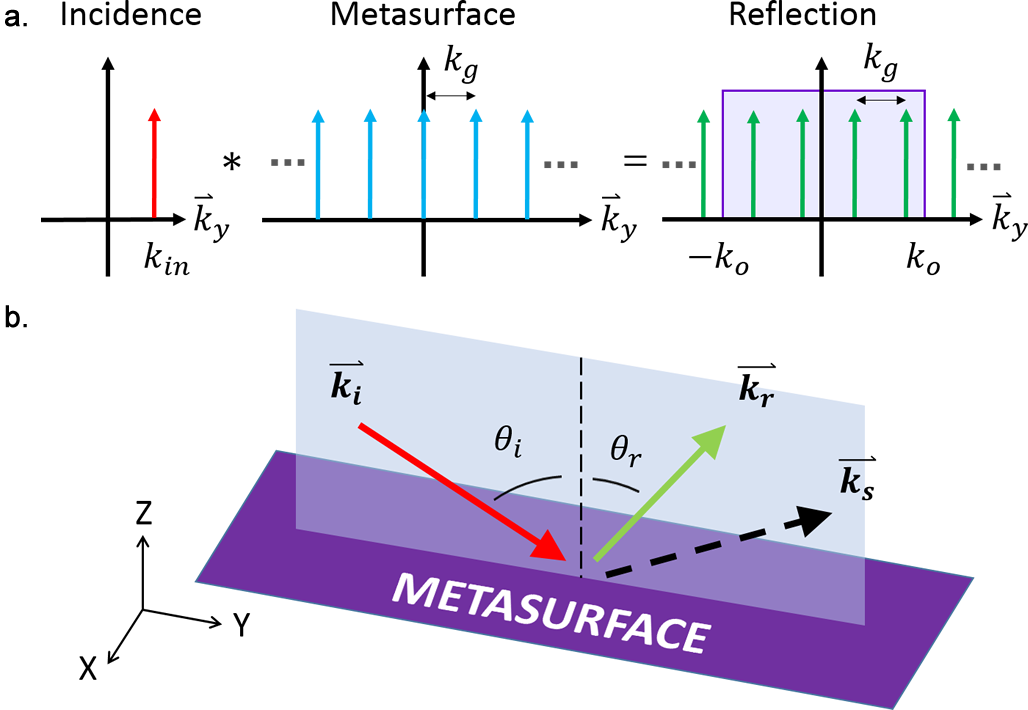}
\caption{\textcolor{blue}{Perfect anomalous reflection with a discretized periodic metasurface. (a) A diagram describing how a periodic metasurface generates diffraction modes. Arrows indicate the existence (not the amplitudes) of diffraction modes; the purple box denotes the regime of propagation waves. $y$ denotes the direction of variation along the metasurface. (b) A schematic diagram of a perfect anomalous reflector, which redirects the incident wave ($\mathbf{k_i}$) into an anomalous direction ($\mathbf{k_r}$), and totally suppresses specular reflection ($\mathbf{k_s}$).}
\label{fig:PeriodicMetasurface}}
\end{figure}

\textcolor{blue}{We begin by describing our theoretical results regarding the discretization of a general periodic metasurface, which includes but is not limited to a perfect anomalous reflector.} The common method for metasurface design is to design a continuous impedance/admittance surface which satisfies the interface boundary condition between a given and a desired electromagnetic wave. Thereafter, the surface is finely discretized and implemented using metasurface elements with corresponding electromagnetic properties. Typical metasurface discretizations range from 8 to 12 cells per wavelength (or in some cases per period) \cite{Smith2011,Pfeiffer2013PRL,Epstein2016NatComm}. These discretizations are usually verified to be of sufficiently fine granularity by yielding results which approximate the theoretical or simulated performance of the continuous counterpart. However, we propose to investigate and implement a metasurface that is as aggressively discretized as possible in regards to the number of elements per period. We shall show that an aggressively discretized metasurface will lead to simple, robust and broadband metasurface implementation. Also, we shall show that the aggressively discretized metasurface has a counterintuitive effect of allowing possibilities unattainable by its continuous counterpart, which, in the case of this paper, enables it to achieve perfect anomalous reflection without explicitly involving a transfer of power through surface or leaky waves.

We employ a diffraction-based perspective to investigate sufficient discretization of a periodic metasurface \cite{AMHW2017GASS}. Fig. \ref{fig:PeriodicMetasurface}a shows the $k$-space operation of a metasurface with period $\Lambda_g$ and period $k_g = 2\pi/\Lambda_g$. Upon the incidence of a plane wave, the periodic metasurface reflects multiple diffraction orders depicted by the location of the arrows. The amplitude and phase of the diffraction orders are dependent on the reflective properties of the metasurface. It is instructive to note that, \textcolor{blue} {while an infinite number of diffraction orders exist in $k$-space,} only a finite number of diffraction orders fall within the propagation range of $k_y \in [-k_0, k_0]$. These represent plane waves that scatter into the far-field, with the propagation angle given by

\begin{equation}
\label{eq:AngularMapping}
\sin \theta_r = \frac{k_y}{k_0} \; \mathrm{for} \left|k_y\right| \leq k_0
\end{equation}

\noindent The other diffraction orders represent evanescent waves that remain within the near-field of the metasurface, but do not contribute to the far-field reflection pattern.
\color{blue}The number of outgoing diffraction modes is given by

\begin{equation}
\label{eq:BoundN}
N = 1 + \lfloor \frac{k_0-k_i}{k_g} \rfloor + \lfloor \frac{k_0+k_i}{k_g} \rfloor.
\end{equation}

\noindent We show in the supplemental material that a metasurface with $N$ independent degrees of freedom is sufficient for tuning the reflection dynamics into each propagation diffraction order. Further, we show that these $N$ degrees of freedom can be met by regenerating the electromagnetic fields at $N$ sample points, equidistant along the length of the period, using an array of $N$ Huygens' sources.

\color{black}

It is of interest to note that, for a class of metasurfaces with sufficiently large spatial frequency such that $k_g \in [k_0, 2k_0)$, there exists only two propagating diffraction orders. This condition is satisfied by many metasurfaces: for example, in retroreflection metasurfaces with $\left| \theta_i \right| = \left| \theta_r \right| \geq 19.5^\circ$ \cite{Hessel1975,Memarian2017,AMHW2017NRSM}. \textcolor{blue}{We shall show in the next sub-section that this also is the case for many perfect anomalous reflectors.} For these cases, the metasurface can be aggressively discretized to having only two elements per grating period  \cite{AMHW2017GASS}. We name the resultant surface a Binary Huygens' Metasurface (BHM). In a previous work we have designed a BHM to perform near-grazing angle retroreflection \cite{AMHW2017NRSM}. We shall proceed to show that a properly designed BHM can also perform perfect anomalous reflection across a wide range of frequencies.

\begin{figure}[t]
\includegraphics[width=87mm]{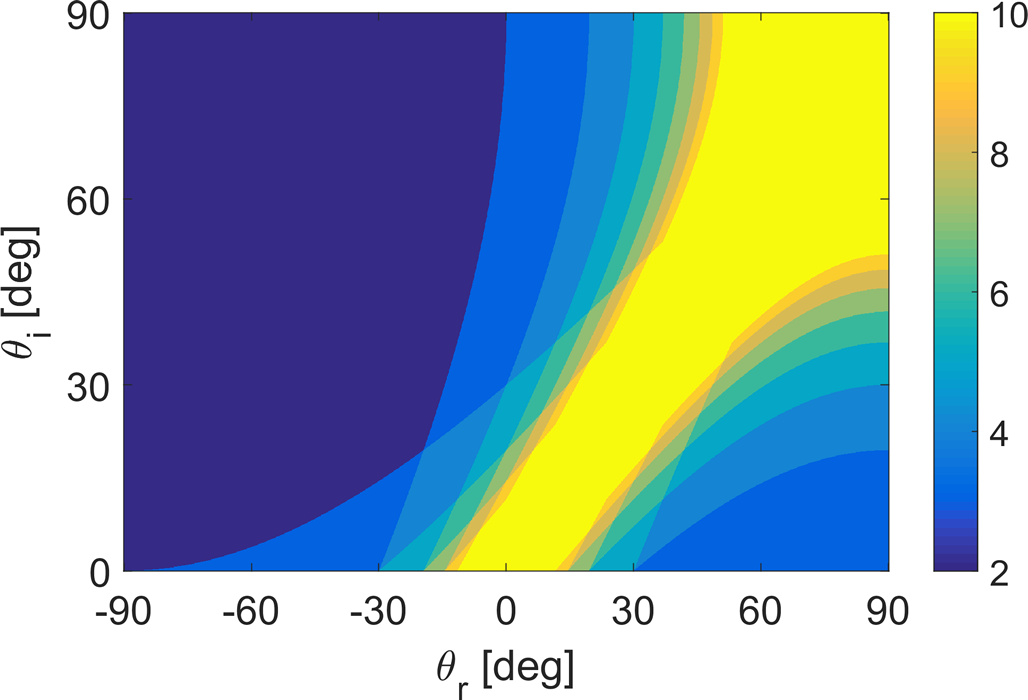}
\caption{\textcolor{blue}{A plot showing $N$ as a function of $\left( \theta_i, \theta_r \right)$}.
\label{fig:N_vs_thetas}}
\end{figure}

\begin{figure*}[t]
\includegraphics[width=120mm]{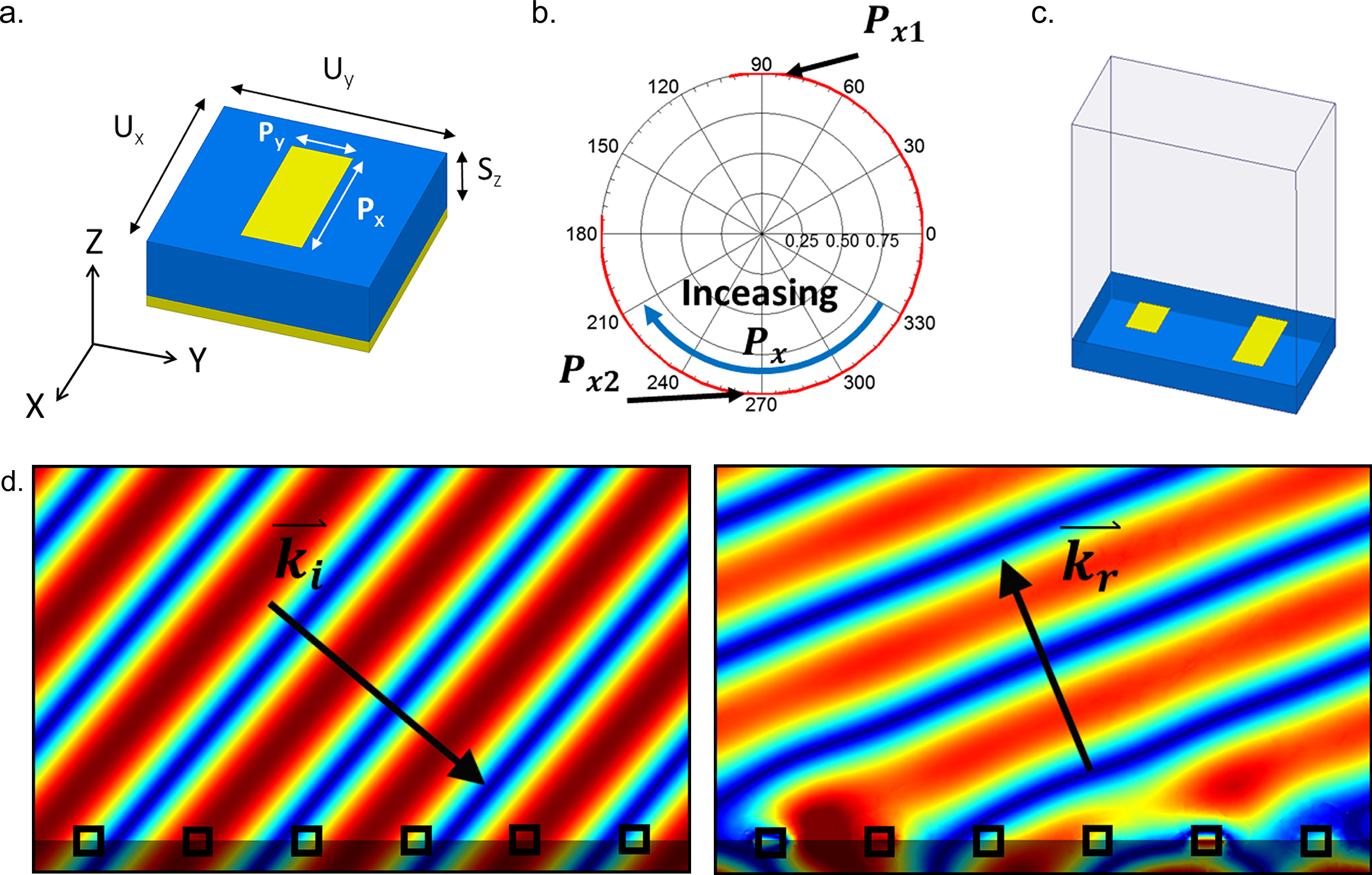}
\caption{Design and simulation of a perfect anomalous reflection metasurface. (a) A metasurface element cell. $U_x=U_y=\Lambda/2$, $S_z = 1.575$ mm, $P_y = 1.5$ mm, $P_x$ is swept to generate the phase shift. (b) The reflection coefficient as $P_x$ is varied. The chosen operation points ($P_{x1}$, $P_{x2}$) are labeled. (c) A schematic of one period of the Binary Huygens' Metasurface. (d) $|\mathbf{E}(\mathbf{r},t)|$ in the plane of incidence, showing the electric field magnitude of the incident (left) and the scattered (right) waves. In both cases, the darkened strip denotes the metasurface, black squares highlight the location of the dipoles.
\label{fig:DesignSimulation}}
\end{figure*}

\subsection{Perfect Anomalous Reflector: Design and Simulation}

\color{blue}
We now concentrate our discussion on a perfect anomalous reflection metasurface. A schematic of a perfect anomalous reflector is shown in Fig. \ref{fig:PeriodicMetasurface}b. The surface accepts an incident wave at $\theta_i$ and anomalously reflects it at an anomalous reflection angle $\theta_r$. Specular reflection ($\mathbf{k_s}$) is totally suppressed. To construct an aggressively discretized perfect anomalous reflection metasurface, one can choose the metasurface period such that $k_g = k_r-k_i=k_0\left(\sin \theta_r - \sin \theta_i\right)$. Substituting this into \eqref{eq:BoundN} yields

 \begin{equation}
\label{eq:BoundN2}
N = 1 + \lfloor \frac{\left(1- \ sin \theta_i \right)}{\left| \sin \theta_r - \sin \theta_i \right|} \rfloor + \lfloor \frac{\left(1+ \ sin \theta_i \right)}{\left| \sin \theta_r - \sin \theta_i \right|} \rfloor.
\end{equation}

Fig. \ref{fig:N_vs_thetas} plots this relationship for $\theta_i \in [0^\circ, 90^\circ]$ and $\theta_r \in [-90^\circ, 90^\circ]$. The sizeable region shaded in dark blue ($N=2$) on the left side of the plot indicates the region for which there exists only two propagating diffraction orders. Hence a binary Huygens' metasurface suffices to perfectly redirect the incident power from the incident to the anomalous reflection angle. Hence for the remaining regions $N > 2$, more metasurface elements are needed for every period. The yellow region ($N \geq 10$) denotes sets of $(\theta_i, \theta_r)$ for which many (ten or more) elements are needed. However, this region also represents `weak' anomalous reflection, in the sense that the reflection angle is only slightly deviated from the incident angle ($\theta_r \approx \theta_i$). Therefore, the period is also correspondingly large, which allows the elements to be spaciously spaced despite the need for many elements within one period.

As a design example demonstrating the Binary Huygens' metasurface, \color{black}
we design our metasurface for the microwave regime at a frequency of 24 GHz. We construct the metasurface on a printed circuit board (PCB) with 1.575 mm thickness and a permittivity of $\varepsilon = 2.2$. We choose as the metasurface element the ground-backed dipole, which, as we have shown in previous works \cite{Kim2014,AMHW2017NRSM,AMHW2017GASS}, can operate as an efficient Huygens' metasurface component. This also exhibits various desirable properties such as low loss, effective phase tuning and a reasonably large tolerance to bandwidth and angular variations. A schematic of the unit cell is shown in Fig. \ref{fig:DesignSimulation}a. We seek to design a metasurface that performs perfect anomalous reflection to convert a TE wave from an incident angle of $50^\circ$ to an anomalously reflected angle of $-22.5^\circ$. A substitution into \eqref{eq:BoundN2} shows that there are only two propagation modes for a metasurface with this periodicity. Hence a Binary Huygens' Metasurface can achieve perfect anomalous reflection into the required direction.

We describe our design process and report full-wave electromagnetic simulation results using the commercial simulator Ansys HFSS. We first characterize the reflection properties of a single element when placed in an infinitely periodic array in the $x$- and $y$- directions. This is achieved by a full-wave Floquet (periodic) simulation. An electromagnetic plane wave, with electric field pointing in the $x$-direction, impinges the element at $\theta_i = 50 ^\circ$, and specular reflection is measured. Fig. \ref{fig:DesignSimulation}b shows the variation in specular reflection as we vary the dipole length $P_x$. We observe that we achieve a near-unity reflection magnitude throughout the sweep, while the reflection phase differs by about $180 ^\circ$ at the operation points $P_{x1} = 2.0$ mm and $P_{x2} = 3.4$ mm. As in \cite{AMHW2017NRSM}, we combine elements which exhibit $180 ^\circ$ difference in phase behaviour to suppress specular reflection and thereby efficiently redirect an incoming wave towards the desired anomalous reflection.

We hence combine the aforementioned elements to form the BHM. When these metasurface elements are placed adjacent to each other, their mutual coupling dynamics cause slight deviations in their reflection properties, which in turn degrade the suppression of specular reflection. To account for this, we sweep $P_{x2}$ from 3.4 mm to 3.8 mm in this new environment, and find that at $P_{x2} = 3.6$ mm, we reestablish destructive interference in the specular direction, and thereby optimize power transfer to the anomalous direction. The resultant Floquet simulation shows that this BHM transfers $99.98\%$ of the incident power from $\theta_i = 50 ^\circ$ to $\theta_r=-22.5 ^\circ$; $0.02\%$ of the power remains as specular reflection. This demonstrates ``perfect'' anomalous reflection to within the accuracy of the simulation, and certainly exceeds the power transfer efficiency as dictated by the impedance mismatch relationship \eqref{eq:ImpedanceMismatch}.

Fig. \ref{fig:DesignSimulation}d shows the incident and scattered waves obtained from the simulation; animations showing the phase progression of both waves are available in the supplemental material that accompanies this paper \cite{SM2}. Both the figure and the animations show that the incident wave is reflected anomalously in a near-perfect manner. In accordance to electromagnetic theory, in a scenario of perfect power transfer from the incident wave into anomalous reflection, the reflected wave features an electric field that differs in amplitude from the incident wave:

\begin{equation}
\label{eq:ReflectedFieldAmp}
\frac { \lvert E_{x,r} \rvert } {\lvert E_{x,i} \rvert} = \sqrt{\frac{\cos \theta_i}{\cos \theta_r}} = 0.8341 \; \mathrm{(for~our~case)}. \\
\end{equation}

Notwithstanding this lower field amplitude, perfect power transfer is achieved because the anomalously reflected wave carries power more efficiently in the $z$-direction.

\subsection{Power Flow Analysis}
\begin{figure}[t]
\includegraphics[width=85mm]{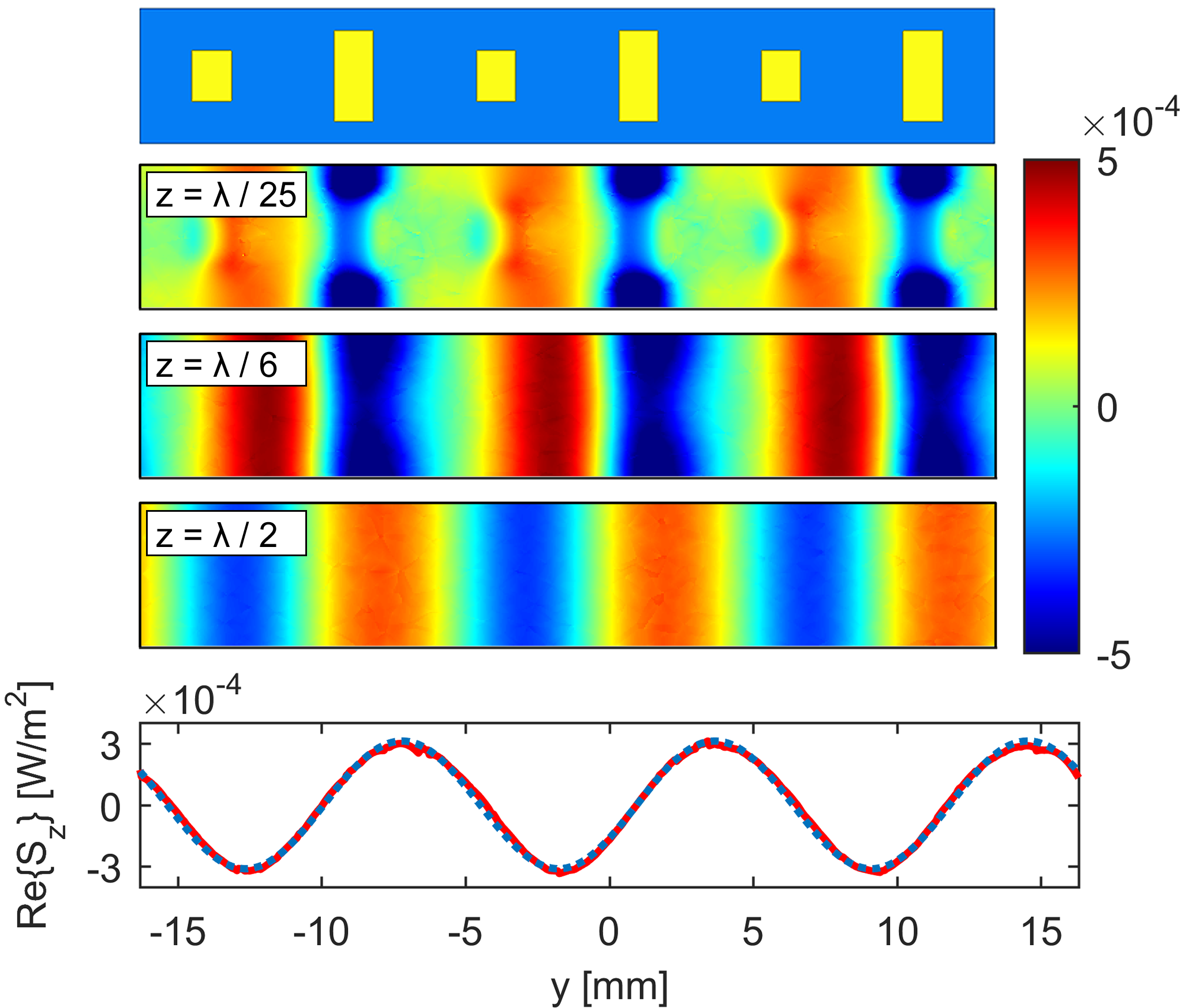}
\caption{Power flow analysis. The top panel shows the metasurface geometry across three spatial periods. The second to fourth panels plot power flow ($\operatorname{Re}\{S_z\}$) across the planes at designated distances above the metasurface. The bottom panel compares $\operatorname{Re}\{S_z(x=0, z=\lambda/2)\}$ (red, solid) against theory (blue, dashed), as calculated by \eqref{eq:PowerFlow}.
\label{fig:PowerFlow}}
\end{figure}

Fig. \ref{fig:PowerFlow} examines the power flow dynamics of the BHM. The top panel of the figure shows the geometry of the metasurface, while the middle three panels show the $z$-directed Poynting vector $\operatorname{Re}\{S_z\}$ (electromagnetic power flow) across three planes, at distances $z = \lambda / 25$, $\lambda / 6$ and $\lambda / 2$ above the metasurface. $\operatorname{Re}\{S_z\} > 0$ implies that power is radiated by the metasurface; conversely, $\operatorname{Re}\{S_z\} < 0$ implies that power is absorbed. In close proximity to the metasurface ($z = \lambda / 25$), we see that the power flow forms an intricate pattern due to the reactive near-field. However, as $z$ increases, the evanescent field diminishes and $\operatorname{Re}\{S_z\}$ approaches a sinusoid. The bottom panel plots the variation at $x=0$ and $z=\lambda/2$ (red, solid). It shows that this variation very well approximates the theoretical sinusoidal variation (blue, dotted) of a perfect anomalous reflection metasurface \cite{Asadchy2016,Estakhri2016}

\begin{equation}
\label{eq:PowerFlow}
\operatorname{Re}\{S_z\} = \frac{\lvert E_{x,i} \rvert ^2}{2 \eta_0} \sqrt{ \frac{\cos \theta_i}{\cos \theta_r}} \left( \cos \theta_r - \cos \theta_i \right) \cos \left( k_g\left(y-\phi\right)\right), \\
\end{equation}

\noindent where $\eta_0$ is the free-space wave impedance and $\phi$ is a constant reflection phase offset. It was previously noted that perfect anomalous reflection required a redistribution of power, such that half the surface is active ($\operatorname{Re}\{S_z\} > 0$) and the other half is lossy ($\operatorname{Re}\{S_z\} < 0$). We have hereby shown that, notwithstanding its apparent simplicity, the BHM successfully accomplishes this necessary redistribution of power through aggressive discretization and the consideration of diffraction modes. We find that this process involves evanescent waves as in \cite{Epstein2016PRL}, but in an implicit manner which simplifies rather than complicates the task of metasurface design and construction.

\begin{figure}[t]
\includegraphics[width=85mm]{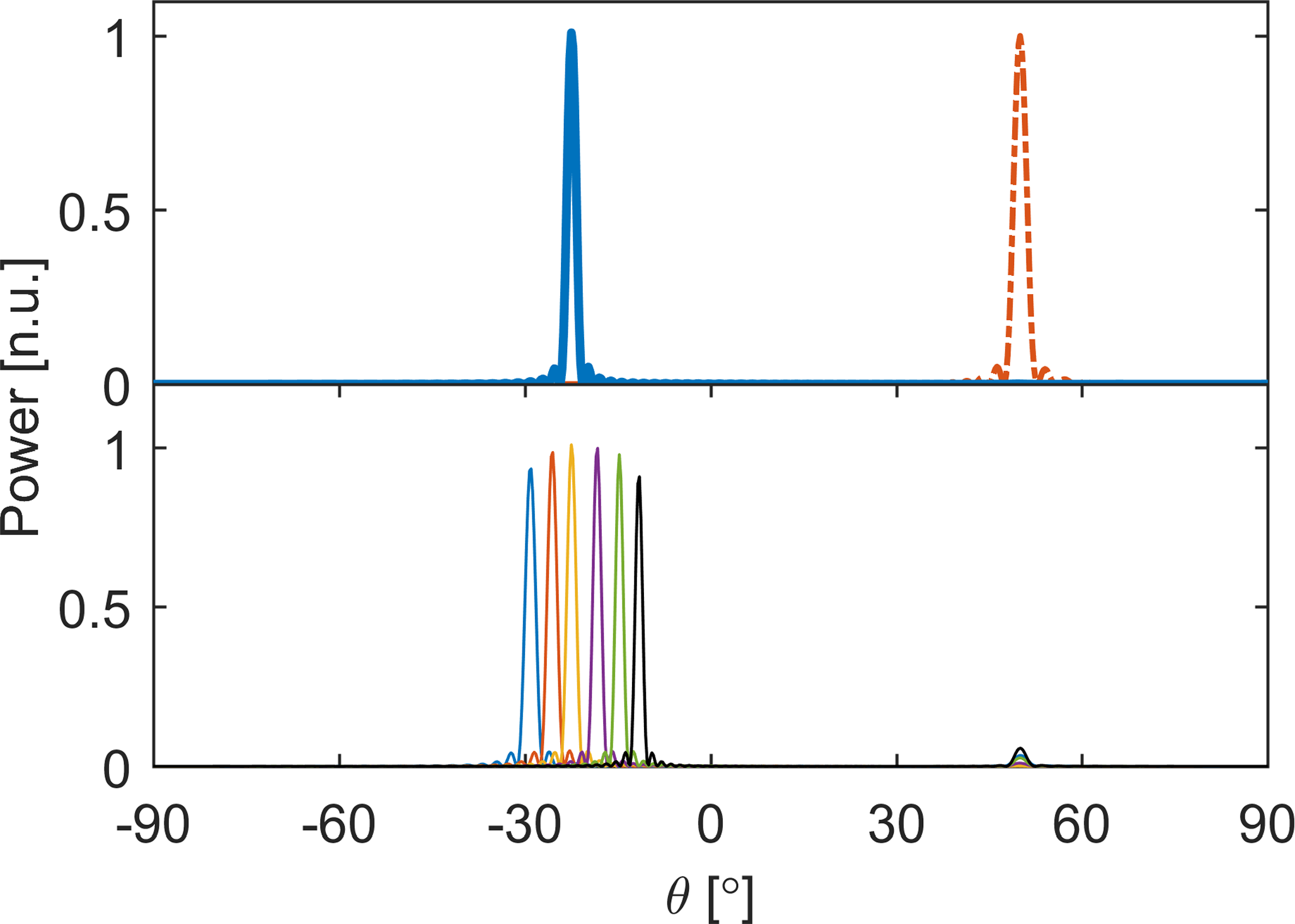}
\caption{Simulated Scattering of a finite metasurface. The top panel compares the normalized scattering magnitude of the BHM (blue, solid) with that of a metallic plate of the same size (red, dashdot). The bottom panel shows the BHM's scattering magnitude over a range of frequencies. Displayed frequencies (from left to right): 22 GHz, 23 GHz, 24 GHz, 25.5 GHz, 27 GHz and 28.5 GHz.
\label{fig:Broadband}}
\end{figure}

\subsection{Finite Surface and Bandwidth Analysis}
We proceed to simulate the reflection properties of the metasurface when it is truncated to a finite size of 78 cells (413.4 mm) in the $y$-direction. Fig. \ref{fig:Broadband} (top panel) shows the reflection characteristics under a 24 GHz plane wave illumination at the designed incident angle. Specular reflection is suppressed to 30 dB (1000 times) below the anomalous reflection power, hence most of the power is anomalously reflected to the desired direction of $\theta_r = -22.5 ^\circ$. To obtain a measurement of power efficiency, the peak of the scattered wave is normalized to the geometric mean of simulated specular reflections for $\theta_{i1} = 50 ^\circ$ and $\theta_{i2} = 22.5 ^\circ$ for a perfect conductor of the same size as the metasurface. (We refer the interest reader to the supplemental material for detailed explanation on our normalization procedure \cite{SM1}. Hence a peak magnitude of 0.03 dB (or $100.7\%$) conveys that the finite-sized metasurface has achieved perfect anomalous reflection to within the accuracy of the simulation.

To illustrate the broadband performance of the metasurface, Fig. \ref{fig:Broadband} (bottom panel) shows the reflection characteristics for a range of frequencies from 22 to 28.5 GHz. Due to the non-resonant nature of the metasurface element, the specular reflection remains significantly suppressed throughout this bandwidth, thus enabling the metasurface to achieve anomalous reflection at a very high ($\geq 90 \%$) efficiency over this bandwidth. Beam squinting --- the variation in beam angle with respect to frequency \cite{Mailloux2005} --- is observed in accordance to antenna array theory. While perfect anomalous reflection is demonstrated at the design frequency, a high-efficiency anomalous reflection of over $90\%$ and significant improvement over the limitation of impedance mismatch is also achieved over this very large fractional bandwidth of $25.7\%$.

\begin{figure*}[!ht]
\includegraphics[width=120mm]{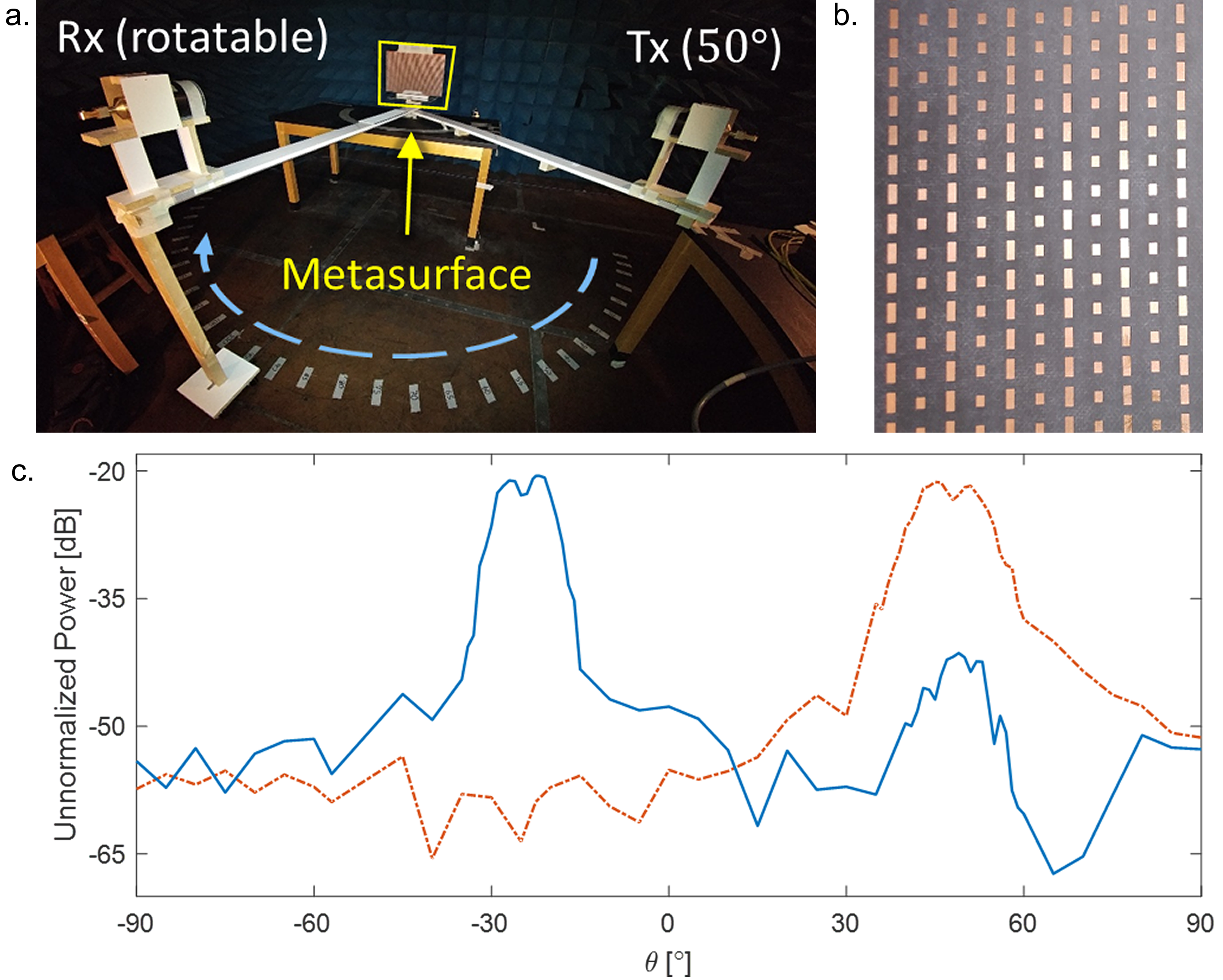}
\caption{Experimental Demonstration. (a) A photo of the experimental apparatus, showing the transmit (Tx) and receive (Rx) antennas and the test surface platform. (b) A close-up of the BHM. (c) Experimentally measured bistatic RCS for the BHM (blue, solid), compared against that for a metallic plate of the same size (red, dashdot).
\label{fig:Experiment}}
\end{figure*}

\subsection{Experimental Demonstration}
We fabricated and experimentally measured anomalous reflection from the Binary Huygens' metasurface. We fabricated the BHM on a Rogers RT/Duroid 5880 laminate board, with 1.575 mm thickness and plated with 1 oz. (35.6 $\mathrm{\mu}$m) copper on both sides. The substrate has a permittivity of $\varepsilon_r = 2.2$ (as in the simulation) and a low dissipation factor of $\tan \delta \in [0.0004, 0.0009]$. Figs. \ref{fig:Experiment}a and \ref{fig:Experiment}b show the fabricated metasurface and the experimental apparatus. A K-band lens antenna emits a vertically- ($x$-) polarized Gaussian beam at $\theta_i = 50 ^\circ$, about 1.5 m from the test surface. A second lens antenna receives the scattered signal, and is swept across the $yz$-plane to measure the test surface's angular scattering profile. To calibrate the system, we first measure the specular reflection strengths of an aluminum test plate with the same size as the Binary Huygens' Metasurface. We find that the power reflected from the aluminum plate at $\theta_{i1} = 50 ^\circ$ and $\theta_{i2} = 22.5 ^\circ$ differ by about 0.5 dB --- which is within the expected error of the experimental apparatus. This implies that the incident Gaussian beam is concentrated at the center of the test surface, such that the power loss due to spillover was negligible. Hence the strength of the anomalous reflection can be found through direct comparison to specular reflection strength for $\theta_i = 50 ^\circ$. Fig. \ref{fig:Experiment}c compares the scattering pattern of the BHM (blue, solid) to that of the metallic test plate (red, dashdot). We observe that the angular spread of the Gaussian beam broadens the reflection features (both by the metallic plate and by the BHM) to a 3 dB (half-power) width of around $10 ^\circ$. Notwithstanding, it is clear that the anomalous reflection surface suppresses specular reflection by about 20 dB, and achieves perfect anomalous reflection to within the sensitivity of the experimental apparatus.

\section{Conclusion}

In this paper we have shown that perfect anomalous reflection is possible with an aggressively discretized metasurface. In particular, we demonstrated that a Binary Huygens' Metasurface, built from ground-backed electric dipoles, can steer an incoming electromagnetic wave at $\theta_i = 50 ^\circ$ to a reflection angle $\theta_r = -22.5 ^\circ$, with perfect efficiency to within the tolerance of full-wave simulation and experimental measurement. In departure to previous proposals, the proposed BHM is surprisingly simple in its single layer structure and large and simple element sizes, which bodes well to the BHM's practical fabrication even at higher frequencies, such as mm-wave, terahertz and beyond. Further, we demonstrate that due to the non-resonant nature of the elements involved, the reported BHM attains wideband performance: though the BHM was designed for operation at 24 GHz, anomalous reflection with greater than $90\%$ efficiency was demonstrated from 22 GHz to 28.5 GHz. Examining the power flow normal to the metasurface reveals that the discretized metasurface implicitly invokes the necessary evanescent waves which facilitate power redistribution from one part of the metasurface to another, as found necessary in \cite{Asadchy2016,Estakhri2016}. This implicit involvement of evanescent waves may also have been the operation mechanism through which traditional anomalous reflection gratings achieved the power redistribution along the reflection surface. In this perspective, the present work reconciles earlier observations of seemingly efficient anomalous reflection in gratings with recent developments in electromagnetic metasurfaces.

During the preparation of this paper, a work \cite{Radi2017} on `meta-gratings' has come to our attention, which also takes the perspective of diffraction modes to achieve high-efficiency wavefront steering. While the aforementioned work numerically investigated wavefront steering using complex single grating elements, the present paper investigates, through both full-wave simulation and experiment, wavefront design with multiple (two or more) simple elements, distributed within a single period of the metasurface. Further, in this work the usage of non-resonant metasurface elements has led to the experimental demonstration of broadband anomalous reflection beyond what has been reported by previous works.

\end{document}



\title{Perfect Anomalous Reflection with a Binary Huygens' Metasurface: Supplemental Material}


\author{Alex M.H. Wong and George V. Eleftheriades}
\email[]{gelefth@ece.utoronto.ca}
\affiliation{The Edward S. Rogers Department of Electrical and Computer Engineering, University of Toronto, Toronto, Canada}


\date{\today}

\maketitle


\section{Maximal Discretization of a Periodic Metasurface}

In this section we investigate the degree to which a periodic metasurface can be discretized. As depicted in Fig. 1 in the main text of the paper, upon plane wave incidence, the wave scattered by a periodic metasurface can be described as a combination of propagating and evanescent modes.

\color{blue}
\begin{equation}
\label{eq:EScat}
 E_{scat}\bigg|_{z=0^+} = \sum_{m} E_m \exp\left(i k_m y\right) \; \mathbf{\hat{x}}. \\
\end{equation}

\color{black}
We note that each mode can be described by the set $\left\{E_m, k_m\right\}$, where $E_m$ is the complex phasor of the $m$'th diffraction order and $k_m = k_i + mk_g$, where $k_i$ and $k_g$ respectively represent the incident wave vector in the $y$-direction and the metasurface spatial frequency. The corresponding magnetic field for each mode is

\begin{equation}
\label{eq:HScat}
\mathbf{H_m} = -\frac{E_m}{\eta_0}\left(\sqrt{k_0^2 - k_m^2} \; \mathbf{\hat{y}} \; + \; k_m \; \mathbf{\hat{z}} \right), \\
\end{equation}

\noindent where $\eta_0$ is the intrinsic impedance of free-space. Since the determination of $\mathbf{H_m}$ does not give extra information or require extra constraints, its mention is omitted hereafter in this section.

We now label by $N$ the number of propagating plane waves emanating from this periodic metasurface, and shift our index from $m$ to $n$, such that the propagation modes are labeled by $n=1$ to $N$. We conjecture that, in order to determine (or control) the set $\left\{E_n\right\}$ for $n=1$ to $N$ by fixing the output electric field for $N$, it is sufficient to control scattered electric field values immediately on top of the metasurface, for $N$ points within the metasurface period. We shall make the practical choice of choosing $N$ equispaced points within the period. We shall label the chosen sample locations $y_p$ and the respective scattered electric fields $a_p$, for $p =1$ to $N$. Now, supposing we know $\left\{a_p, y_p, k_n\right\}$, we can find $E_n$ through a Fourier transform

\begin{equation}
\label{eq:FT}
E_n = C \sum_{p=1}^N a_p \exp(-i k_n y_p), \\
\end{equation}

\noindent where $C$ is a power normalization constant. Writing in a matrix form, we have

\begin{equation}
\label{eq:FTMatrix}
\vec{E} = \underline{\underline{M}}\vec{A}, \; \mathrm{where~} \vec{E} =
\begin{bmatrix} E_1 \\ E_2 \\ \vdots \\ E_N \end{bmatrix},
\underline{\underline{M}} = C
\begin{bmatrix}
  e^{-ik_1 y_1}  & e^{-ik_1 y_2}    & \dots & e^{-ik_1 y_N} \\
  e^{-ik_2 y_1}  & e^{-ik_2 y_2}    & \dots & e^{-ik_2 y_N} \\
  \vdots            & \vdots                & \ddots    & \vdots \\
  e^{-ik_N y_1}  & e^{-ik_N y_2}    & \dots & e^{-ik_N y_N} \\
\end{bmatrix},
\mathrm{~and~} \vec{A} =
\begin{bmatrix} a_1 \\ a_2 \\ \vdots \\ a_N \end{bmatrix} . \\
\end{equation}

\noindent This shows that an $N$-fold discretization within the metasurface period sufficiently determines the $N$ diffraction orders which propagate into the far-field. Conversely, since the $N \times N$ matrix  $\underline{\underline{M}}$ is invertible ($\underline{\underline{M}}^{-1}$ is the inverse Fourier transform matrix), one can also fully determine, by inverting \eqref{eq:FTMatrix}, the necessary values which generate a desired superposition of scattered waves.

\color{blue}
While we have heretofore focused on the propagating diffraction modes which affect the far-field reflection of the metasurface, the Fourier formulation of \eqref{eq:FT} is general to all modes, hence it can also be used to determine evanescent diffraction modes which are generated and remain within the near-field of a metasurface. A straightforward application of \eqref{eq:FT}, or more generally the concept of discrete-space Fourier series \cite{Oppenheim1996}, will lead to the conclusion that the complex amplitude of the diffraction modes are actually periodic, with $\left\{E_1, ..., E_n \right\}$ forming one complete period in $k$-space. However, the amplitude of excitations for strongly evanescent diffraction modes will be quelled, due to the finite bandwidth of the spectral profile of the individual metasurface element, which is multiplied onto any ``array-factor'' spectral coefficients like those derived in \eqref{eq:FT}. In summary, the formulation hereby described enables one to derive both propagating and evanescent diffraction modes which arise in the metasurface scattering process. A refinement that multiplies \eqref{eq:FT} with the spectral profile of the metasurface element will improve the accuracy of the diffraction modes, especially in the evanescent region, but its effects is of less significance in the propagation mode regime. Therefore it is excluded for simplicity, in similarity to treatments in antenna array design.

\newpage

\color{black}

\section{Power Normalization for a Finite Anomalous Reflection Metasurface}

In general, the amount of power that couples from an electromagnetic plane wave into an antenna (or conversely, the amount of power radiated by the antenna and emanates as a plane wave) is strongly dependent on the projected size of the antenna. The projected size of a finite surface is the projection of the surface onto a plane normal to an incoming plane wave. As depicted in Fig. S2, for a plane wave incidence at an angle $\theta$ with respect to broadside, the projected aperture of a rectangular surface with area $A$ is

\begin{equation}
\label{eq:PhysApt}
A_p = A \cos \theta
\end{equation}

A uniformly excited aperture has a directivity that relates linearly with $A_p$. That is to say, the amount of power it receives from an illuminating plane wave, as well as the amount of power a properly phased aperture radiates in the $\theta$ direction, both individually depend on $A_p$. It is hence understood why the monostatic radar cross section (RCS) of an object is related to the square of its effective aperture area \cite{Mahafza2005,Bird2005}. The same applies for the specular reflection of an electrically large rectangular metallic plate: both the amount of power which illuminates the plate and the amount of power which reradiates into the specular direction depend linearly on $A_p$. Hence the specularly reflected power from a metallic plate is given by

\begin{figure}[b]
\includegraphics[width=86mm]{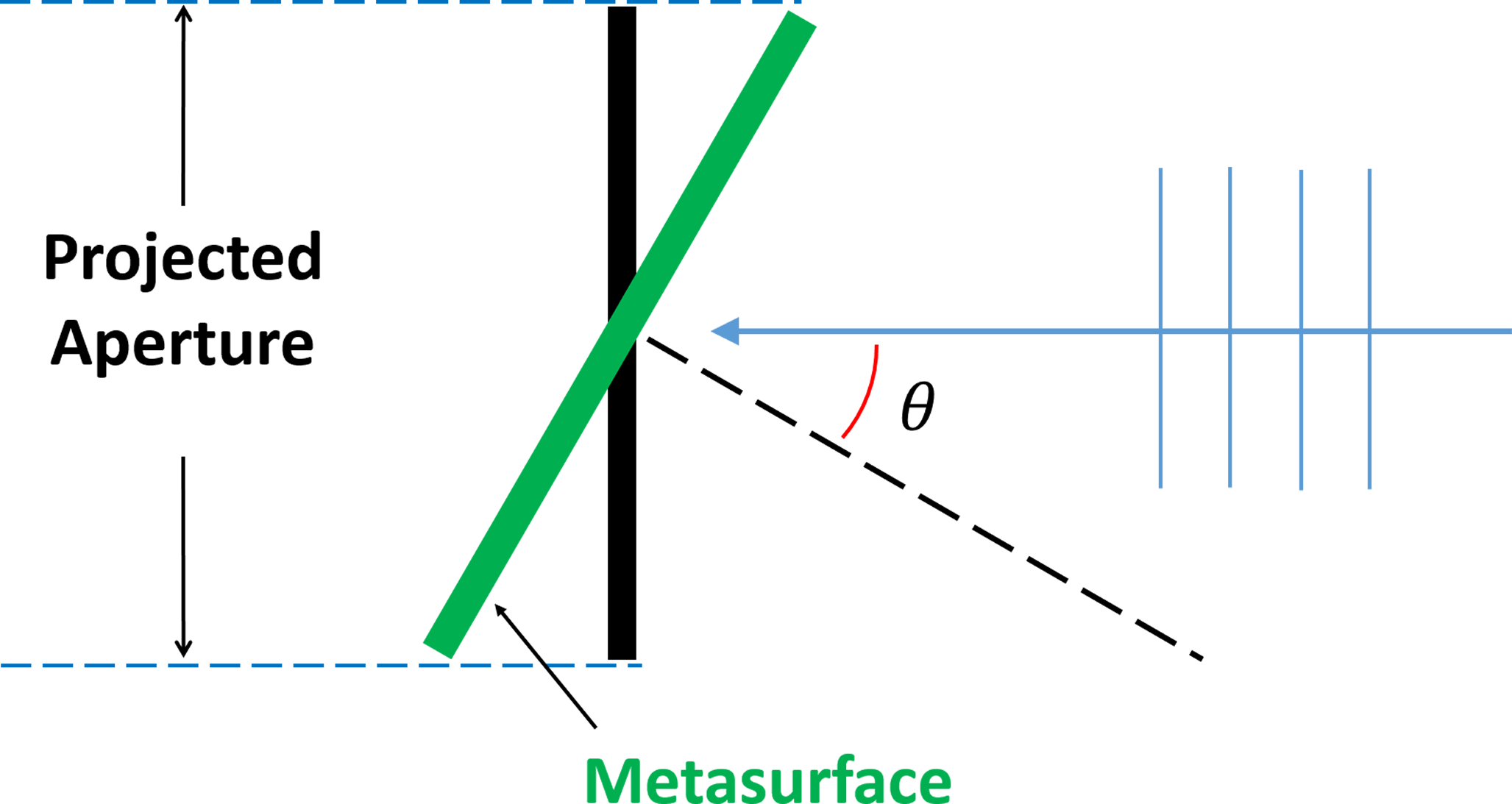}
\caption{A diagram illustrating the projected aperture of a rectangular metasurface, as a function of the illumination angle.
\label{fig:ProjectedAperture}}
\end{figure}

\newpage

\begin{equation}
\label{eq:MetalPlateRefl}
P_{metal}(\theta) = P_{0,metal} \cos^2 \theta, \\
\end{equation}

\noindent where $P_{0,metal}$ is the reflected power at normal incidence.

An anomalously reflected metasurface features different incidence and reflection angles, which slightly complicates the task of characterizing its efficiency in simulation and experiment, specifically when the surface is not large enough to capture most of the initial illumination. One reasonable way to determine the anomalous reflection efficiency is

\begin{equation}
\label{eq:AnomalousRefl}
P_{ar}(\theta_i, \theta_r) = \eta_{ar} P_{0,metal} \cos \theta_i \cos \theta_r. \\
\end{equation}

\noindent Here $P_{ar}(\theta_i, \theta_r)$ and $\eta_{ar}$ represent the anomalously reflected power and power efficiency respectively. Rearranging \eqref{eq:MetalPlateRefl} and \eqref{eq:AnomalousRefl} gives

\begin{equation}
\label{eq:AnomalousReflEff}
\eta_{ar} = \frac{P_{ar}(\theta_i, \theta_r)}{\sqrt{P_{metal}(\theta_i) P_{metal}(\theta_r)}}. \\
\end{equation}

This allows one to find the power conversion efficiency of the anomalous reflection metasurface in terms of simulated and/or measured reflection reflection powers of the metasurface and a same-sized metallic plate.